\documentclass[runningheads]{svmult}
\usepackage{makeidx}   % allows index generation
\usepackage{graphicx}  % standard LaTeX graphics tool
\usepackage{subeqnar}  % subnumbers individual equations
\usepackage{multicol}  % used for the two-column index
\usepackage{cropmark} % cropmarks for pages without
\usepackage{physprbb}  % centered layout of diverse elements, etc.
\begin{document}
\title*{Dark Energy in Extra Dimensions and 
\newline String Theory: Consistency Conditions\footnote{
Talk given at ``Third International Symposium on Dark Matter in
Astro and Particle Physics'', dark2000, Heidelberg, July 2000 and
SUSY 2K, CERN, Geneva, June 2000. To appear in the proceedings
of dark2000.}}
\toctitle{Dark Energy in Extra Dimensions and
\protect\newline String Theory: Consistency Conditions}
% allows explicit linebreak for the table of content
%
%
\titlerunning{Dark Energy in Extra Dimensions and String Theory}
% allows abbreviation of title, if the full title is too long
% to fit in the running head
%
\author{Hans Peter Nilles\inst{1}}
%\and Roger Temam\inst{2}
%\and Jeffrey Dean\inst{2}
%\and David Grove\inst{1}
%\and Craig Chambers\inst{2}
%\and Kim~B.~Bruce\inst{2}
%\and Elsa Bertino\inst{1}}
%
\authorrunning{Hans Peter Nilles}
% if there are more than two authors,
% please abbreviate author list for running head
%
%
\institute{Physikalisches Institut, Universit\"at Bonn,
           Nussallee 12, 53115 Bonn, Germany}

%\and Universit\'{e} de Paris-Sud,
%     Laboratoire d'Analyse Num\'{e}rique,
%     B\^{a}timent 425,\\
%     F-91405 Orsay Cedex, France}

\maketitle              % typesets the title of the contribution

\begin{abstract}

The smallness of the cosmological constant is one of the basic
problems in particle physics and cosmology. Various attempts
have been made to explain this mystery, but no satisfactory
solution has been found yet. The appearance of extra dimensions 
in the framework of brane world systems seems to provide 
some new ideas to address this problem form a different
point of view. We shall discuss some of these new approaches
and see whether or not they lead to an improvement of the
situation. We shall conclude that we are still far from a
solution of the problem. 

\end{abstract}

\section{Introduction}

We know that the cosmological constant is much smaller than one
would naively expect. This led to the belief that a natural
approach to this problem would be a mechanism that explains
a vanishing value of this vacuum energy. While cosmological
observations\cite{Perlmutter:1999np,Garnavich:1998nb} seem
to be consistent with a nonzero value of the
cosmological constant, still the small value obtained lacks
a satisfactory explanation other than just being the result of
a mere fine-tuning of the parameters. 

Recently new theoretical ideas in extra
dimensions have been put forward to attack this
problem. In the present talk I shall elaborate on work done in
collaboration with Stefan F\"orste, 
Zygmunt Lalak and St\'ephane
Lavignac\cite{Forste:2000ps,Forste:2000ft,Forste:2000ge}, 
where the problem
of fine-tuning has been analyzed in the framework of models
with extra dimensions that have attracted some attention recently.

One of the most outstanding open problems in quantum field theory is it
to find an explanation for the 
stability of the observed value of the cosmological
constant in the presence of radiative corrections.
As we will see below (and as has been discussed in
several review
articles\cite{Weinberg:1989cp,Witten:2000zk,Binetruy:2000mh}) a
simple quantum field theoretic estimate provides naturally a
cosmological constant which is at least 60 orders of magnitude to
large. Quantum fluctuations create a vacuum energy which in turn
curves the space much stronger than it is observed. Hence, the
classical vacuum energy needs to be adjusted in a very accurate way in
order to cancel the contributions from quantum effects. 
This would require a fine-tuning of the fundamental parameters  
of the theory to an
accuracy of at least 60 digits. 
From the theoretical point of view we consider this as
a rather unsatisfactory situation and would like to analyze
alternatives leading to the observed
cosmological constant in a more natural way. In this talk we will
focus on brane world scenarios and how they might modify the above
mentioned problem. In brane worlds the observed matter is confined to
live on a hypersurface of some higher dimensional space, whereas
gravity and possibly also some other fields can propagate in all
dimensions. This may give some alternative point of view 
concerning the
cosmological constant since the vacuum energy generated by quantum
fluctuations of fields living on the brane may not curve the brane
itself but instead the space transverse to it. The idea of brane
worlds dates back to \cite{Rubakov:1983bz,Rubakov:1983bb,Akama:1982jy}.
A concrete realization can be found in the context of string theory
where matter is naturally confined to live on
D-branes \cite{Polchinski:1995mt} or orbifold fixed
planes \cite{Horava:1996qa}. More recently there has been renewed
interest in addressing the problem of the cosmological constant within
brane worlds, for an (incomplete) list of references see
[\ref{braneworlda}--\ref{braneworlde}] and references
therein/thereof.  

The talk will be organized as follows. First, we will recall the
cosmological constant problem as it appears in ordinary four
dimensional quantum field theory. We shall then elaborate on
some of the past (four-dimensional) attempts to solve the problem.
Subsequently the general set-up of brane worlds
will be presented. Particular emphasis will be put on
a consistency condition (sometimes also called a sum rule)
for warped compactifications that has been overlooked in various attempts
to address the problem of the cosmological constant and which is a
crucial tool to understand the issue of fine-tunings in the brane
world scenario. 
Then we will study how fine-tunings 
appear in order to achieve a vanishing cosmological constant in the
Randall Sundrum model \cite{Randall:1999ee,Randall:1999vf}. 
We shall argue that a similar fine-tuning 
is needed in the set-up  presented
in \cite{Arkani-Hamed:2000eg,Kachru:2000hf}  once the
singularity is resolved. Finally, we elaborate on the issue of the
existence of nearby curved solutions and we will argue that it is this
questions that has to be addressed if one wants to understand the
small value of the cosmological constant.

\section{The problem}

%\vskip0.5cm
%\noindent
The observational bound on the cosmological constant is
\begin{equation}\label{observ}
\lambda M_{Pl}^2 \leq 10^{-120} \left( M_{Pl}\right)^4 
\end{equation}
where $M_{Pl}$ is the Planck mass (of about $10^{19}$ GeV) and the
formula has been written in such a way that the quantity appearing on the left
hand side corresponds to the vacuum energy density. This is a very 
small quantity once one admits the possiblilty of the Planck
scale as the fundamantal scale of physics. 
Even in the particle physics standard model of weak, strong  
and electromagnetic interactions one would expect a tree level
contribution to the vacuum energy of order of
several hundred GeV taking into account the scalar potential
that leads to electroweak symmetry breaking.
Moreover, in quantum field
theory we expect additional contributions from perturbative
corrections, e.g. at one loop
\begin{equation}\label{qft}
\lambda M_{Pl}^2 = \lambda_0 M_{pl}^2 + \left(\mbox{UV-cutoff}\right)^4
  Str\left( \mbox{\bf 1}\right)
\end{equation}
in addition to
$\lambda_0$ the bare (tree level) value of the cosmological
constant which can in principle be chosen by hand. 
The supertrace in (\ref{qft}) is
to be taken over degrees of freedom which are light compared to the
scale set by the UV-cutoff. Comparison of (\ref{observ}) with
(\ref{qft}) shows that one needs to fine-tune 120 digits in
$\lambda_0 M_{Pl}^2$ such that it cancels the one-loop contributions
with the necessary accuracy. Supersymmetry could ease this problem
of radiative corrections (for a review see\cite{Nilles:1984ge}).
If one believes that the world is supersymmetric 
above the TeV scale one would still need to adjust 60
digits. Instead of adjusting input parameters of the theory to such a
high accuracy in order to achieve agreement with observations one would
prefer to get (\ref{observ}) as a prediction or at least as a natural
result of the theory (in which, for example,  only a few digits need
to be tuned, if at all). 

This is the situation within the framework of four-dimensional
quantum field theories.
The above discussion might be modified in a brane world
setup which we will discuss 
in this lecture. We should however mention already at
this point that ``modification'' does not 
necessarily imply an improvement of the
situation. Before we get into this discussion let us first 
recall some attempts to solve the problem in the 
four-dimensional framework.

\section{Possible solutions?}

A starting point for a natural solution would be 
a symmetry that forbids a cosmological constant. In fact,
symmetries that could achieve this do exist: 
e.g. supersymmetry and conformal
symmetry. Unfortunately these symmetries are badly broken
in nature at a level of at least a few hundred GeV and
therefore the problem remains. Still one might think 
that the presence of such a symmetry would be a first step 
in the right direction. 

A second possible solution could be a dynamical mechanism to
relax the cosmological constant. Such a mechanism could be
quite similar to the axion mechanism that relaxes the
value of the $\theta$ parameter in 
quantum chromodynamics (QCD). 
This mechanism needs a new ingredient, a propagating field that
adjusts is vacuum expectation value dynamically. For a review of
these questions see \cite{Weinberg:1989cp,Nilles:1998uy}.
 In string
theory the so-called ``sliding dilaton'' could play this role as has been
argued in \cite{Derendinger:1985kk,Dine:1985rz}. In all
these cases, however, one would then expect the existence
of an extremely light scalar degree of freedom which 
would  lead to  new fifth force that probably should not 
have escaped our detection.

Other attempts to understand the value of the vacuum energy
have used the anthropic principle in one of its various forms.
For a review see \cite{Weinberg:1989cp}.

Given the present situation it is fair to say that we do not
have yet a satisfactory solution of the problem of
the cosmological constant, at least in the framework of
four-dimensional string and quantum field theories.
Could this be better in a higher dimensional world?
For an alternative way to address the problem in less than four
space-time dimensions see \cite{Witten:2000zk}.

We should keep in mind, however, that the problem of
the cosmological constant is just a problem of fine-tuning
the parameters of the theory in a very special way. We now want 
to see whether this can be avoided in a higher dimensional
set-up.

\section{What about extra dimensions $d>4$?}

In the so-called ``brane world scenario''
matter fields (quarks and leptons, gauge bosons, Higgs bosons) are
supposed to be confined to live on a hyper surface (the brane) in a
higher dimensional space, whereas gravity and possibly also some
additional fields can propagate also in directions transverse to the
brane. Such a picture of the universe is motivated by recent
developments in (open) string theory \cite{Polchinski:1995mt} 
and heterotic M-theory \cite{Horava:1996qa,Horava:1996ma}.
Since gravitational interactions are much weaker than the other
known interactions, the size of the additional dimension is much
less constrained by observations than in usual
compactifications. In fact, the size of the additional dimensions
might be directly correlated to the strength of four-dimensional
gravitational interactions\cite{Witten:1996mz}.
Looking for example on product compactifications of
type I string theory it has been noted that it is possible to push the
string scale down to the TeV range when one allows at least two of the
compactified dimensions to be ``extra large'' (i.e.\ up to a $\mu
m$)\cite{Antoniadis:1998ig}. 

A first look at the question of the cosmological constant does
not look very promising. The naive expectation would be that
the cosmological constant in the extra (bulk) dimensions
$\Lambda_B$ and that on the brane, the brane tension $T$, should
vanish separately. We would then essentially have the same
situation as in the four-dimensional case, with the additional
problem to explain why also $\Lambda_B$ has to vanish. The
known mechanism of a sliding field \cite{Derendinger:1985kk,Dine:1985rz}
can be carried over to this 
case \cite{Horava:1996vs,Nilles:1997cm,Nilles:1998sx,Lukas:1998rb},
but does not shed any new light on the question of
the cosmological constant. 

A closer inspection of the situation reveals the
novel  possibility to have a flat brane even in the presence
of a nonzero tension $T$. For a consistent picture, however, 
here one also has to require a non-zero bulk cosmological constant
$\Lambda_B$ that compensates the vacuum energy (tension) of
the brane. In some way this corresponds to a picture where the
vacuum energy of the brane does not lead to a curvature on
the brane itself, but curves transverse space and leaves the
brane flat. Curvature of the brane can flow off to the
bulk, a mechanism that is sometimes called ``self-tuning''. 

For such a mechanism to appear we need to consider so-called
warped compactifications where brane and transverse space
are not just a direct product. We shall see that in this case
we can have flat branes embedded in higher dimensional
anti de Sitter space, provided certain consistency conditions
have been fulfilled.

In the following we will be considering the special
case that the brane is 1+3 dimensional and we have one additional
direction called $y$. Then the ansatz for the five dimensional metric
is in general ($M,N = 0,\ldots ,4$ and $\mu,\nu =0, \ldots 3$)
\begin{equation}\label{ansatz0}
ds^2 \equiv G_{MN}dx^Mdx^N = e^{2A\left(
    y\right)}\tilde{g}_{\mu\nu}dx^\mu dx^\nu + dy^2   
\end{equation}
where the brane will be localized at some $y$.
We split
\begin{equation}
\tilde{g}_{\mu\nu} = \bar{g}_{\mu\nu} + h_{\mu\nu}
\end{equation}
into a vacuum value $\bar{g}_{\mu\nu}$ and fluctuations around it
$h_{\mu\nu}$. For the vacuum value we will be interested in maximally
symmetric spaces, i.e. Minkowski space ($M_4$), de Sitter space
($dS_4$), or anti de 
Sitter space ($adS_4$). In particular, we chose coordinates such that
\begin{equation}\label{ansatz}
\bar{g}_{\mu\nu} =
\left\{\begin{array}{lll}
diag\left(-1,1,1,1\right) & \mbox{for:} & M_4\\
diag\left(
  -1,e^{2\sqrt{\bar{\Lambda}}t},e^{2\sqrt{\bar{\Lambda}}t},
      e^{2\sqrt{\bar{\Lambda}}t}\right)  & \mbox{for:} & dS_4\\
diag\left(-e^{2\sqrt{-\bar{\Lambda}}x^3},e^{2\sqrt{-\bar{\Lambda}}x^3},
  e^{2\sqrt{-\bar{\Lambda}}x^3}, 1\right) & \mbox{for:} & adS_4
\end{array}\right.
\end{equation}
That means we are looking for 5d spaces which are foliated with
maximally symmetric four dimensional slices.
Throughout this talk, the five dimensional action will be of the form
\begin{equation}\label{action}
S_5 = \int d^5 x\sqrt{-G}\left[ R-\frac{4}{3}\left(
    \partial\phi\right)^2 - V\left(\phi\right)\right] - \sum_i\int
    d^5x \sqrt{-g} f_i\left( \phi\right)\delta\left( y-y_i\right) .
\end{equation}
We allow for situations where apart from the graviton also a scalar
$\phi$ propagates in the bulk. The positions of the branes involved
are at $y_i$. With lower case $g$ we denote the induced metric on the
brane which for our ansatz is simply
\begin{equation}
g_{\mu\nu} = G_{MN} \delta^M _\mu \delta^N _\nu .
\end{equation}
The corresponding equations of motion read
\begin{eqnarray}
\sqrt{-G}\left[ R_{MN} -\frac{1}{2}G_{MN}R -\frac{4}{3} \partial_M\phi
  \partial_N\phi +\frac{2}{3}\left(\partial\phi\right)^2 G_{MN}
  +\frac{1}{2} V\left(\phi\right) G_{MN}\right] & &\nonumber \\
  \;\;\; +\frac{1}{2}\sqrt{-g}\sum_i f_i
  \delta\left(y-y_i\right)g_{\mu\nu}\delta^\mu _M\delta^{\nu}_{N} = 0
  & &,
\end{eqnarray}
\begin{equation}
-\frac{\partial V}{\partial \phi}\sqrt{-G}
 +\frac{8}{3}\partial_M\left( \sqrt{-G}G^{MN}\partial_N\phi\right) -
 \sqrt{-g} \sum_i \delta\left( y- y_i\right)\partial_\phi
 f_i\left(\phi\right) = 0 .
\end{equation}
After integrating over the fifth coordinate in (\ref{action}) one
obtains a four dimensional  effective theory. In particular, the
gravity part will be of the form  
\begin{equation}\label{effact}
S_{4,grav} = M_{Pl}^2 \int d^4 x\sqrt{-\tilde{g}}\left(\tilde{R}
  -\lambda\right) , 
\end{equation}
where $\tilde{R}$ is the 4d scalar curvature computed from
$\tilde{g}$. The effective Planck mass is $M_{Pl}^2 = \int dy e^{2A}$,
(note that we put the five dimensional Planck mass to one).
Now, for consistency the ansatz (\ref{ansatz}) should be a stationary
point of (\ref{effact}). This leads to the requirement $\lambda =
6\bar{\Lambda}$. Finally, the on-shell values of the 4d effective
action should be equal to the 5-dimensional one. This results in the
consistency condition\cite{Forste:2000ft} (see also\cite{Ellwanger:2000pq}),
\begin{equation}\label{consistent}
\frac{\left< S_5\right>}{\int d^4 x} = 6\bar{\Lambda}M_{Pl}^2 .  
\end{equation} 
It has to be fulfilled for all consistent solutions of the
Einstein equations, independently whether the branes are flat or
curved.
Especially for foliations with Poincare
invariant slices the vacuum energy $\bar{\Lambda}$ should vanish.
Curved solutions would require a corresponding nonzero value of
$\bar{\Lambda}$.
It is this adjustment of the parameters that replaces the traditional
four-dimensional fine-tuning in the brane world picture.

\section{A toy example: the Randall-Sundrum set-up}

As a warm-up example for a warped compactification we want to study the
model presented in \cite{Randall:1999ee}. There is no bulk scalar in
that model. Therefore, we put $\phi = const$ in
(\ref{action}). Moreover, we plug in 
\begin{equation}
V\left( \phi\right) = - \Lambda_B \,\,\, ,\,\,\, f_1 = T_1\,\,\, ,
\,\,\, f_2 = T_2 ,
\end{equation}
where $\Lambda_B$, $T_1$ and $T_2$ are constants. There will be two
branes: one at $y=0$ and a second one at $y=y_0$. Denoting with a
prime a derivative with respect to $y$ the $yy$-component of the
Einstein equation gives
\begin{equation}\label{einst}
6\left( A^\prime\right)^2 = -\frac{\Lambda_B}{4}
\end{equation}
Following \cite{Randall:1999ee} we are looking for solutions being
symmetric under $y \rightarrow -y$ and periodic under $y\rightarrow y
+2 y_0$. The solution to (\ref{einst}) is
\begin{equation}\label{rssol}
A= -\left| y\right| \sqrt{ - \frac{\Lambda_B}{24}} ,
\end{equation}
where $\left| y\right| $ denotes the familiar modulus function for
$-y_0 < y < y_0$ and the periodic continuation if $y$ is outside that
interval. The remaining equation to be solved corresponds to the
$\mu\nu$ components of the Einstein equation,
\begin{equation}\label{einsta}
3A^{\prime\prime} = -\frac{T_1}{4}\delta\left( y\right) -
\frac{T_2}{4}\delta\left( y- y_0\right) .
\end{equation}
This equation is solved automatically by (\ref{rssol}) as long as $y$
is neither $0$ nor $y_0$. 
Integrating equation (\ref{einsta}) from $-\epsilon$ to $\epsilon$,
relates the brane tension $T_1$ to the bulk cosmological constant
$\Lambda_B$, 
\begin{equation}\label{ft1}
T_1 = \sqrt{-24\Lambda_B}.
\end{equation}
Integrating around $y_0$ gives
\begin{equation}\label{ft2}
T_2 = -\sqrt{-24\Lambda_B}.
\end{equation}
These relations arise due to $\bar{\Lambda}=0$ in the ansatz and can
be viewed as fine-tuning conditions for the effective cosmological
constant ($\lambda$ in (\ref{effact}))\cite{DeWolfe:2000cp}. Indeed,
one finds that the consistency condition (\ref{consistent}) is
satisfied only when (\ref{rssol}) together with both fine-tuning
conditions (\ref{ft1}) and (\ref{ft2}) are imposed. Since the brane
tension $T_i$ corresponds to the vacuum energy of matter living in the
corresponding brane, the amount of fine-tuning contained in
(\ref{ft1}), (\ref{ft2}) is of the same order as needed in ordinary 4d
quantum field theory discussed in the beginning of this talk. 

Next we have to address 
the important question: What happens if the fine-tunings do
not hold? Does this necessarily lead to disaster or do solutions
exist also in this case.
Indeed it has been shown in \cite{DeWolfe:2000cp} that 
in that case
solutions exist, however with $\bar{\Lambda}\not= 0$. 
This
closes the argument of interpreting conditions (\ref{ft1}) and
(\ref{ft2}) as fine-tunings of the cosmological constant. It also
emphasizes the new problem with the adjustment of the
cosmological constant on the brane: how to select the flat
solution instead of these ``nearby'' curved solutions that are 
continuously connected in parameter (moduli) space.

\section{Going back to $\Lambda_B =0$: does it make sense?}

Thus the generic higher dimensional set-up considers nonzero
values of brane tensions and the bulk cosmological constant.
A fine tuning is needed to arrive at a flat brane with
vanishing cosmological constant.

Recently an attempt has been made to study the situation
with $\Lambda_B=0$.
We will focus on a `` solution'' discussed
in \cite{Arkani-Hamed:2000eg,Kachru:2000hf} (solution II of the second
reference). In this model there is a bulk scalar without a bulk
potential 
\begin{equation}
V\left( \phi\right) =0 .
\end{equation}
In addition we put one brane at $y=0$, and a bulk scalar with a very
specific coupling to
the brane via
\begin{equation}\label{coupling}
f_0 \left(\phi\right) = T_0 e^{b\phi}\,\,\, ,\,\,\,\mbox{with:}\,\,\, b=\mp
\frac{4}{3} .
\end{equation}
Observe that this model already assumes fine-tuned values 
$\Lambda_B$ and $b$ which would have to be explained. We now
make the same warped ansatz (\ref{ansatz0}) as before. If again we
assume $\bar{\Lambda}=0$
in (\ref{ansatz}), the  bulk equations seem to be solved by $A^\prime
=\pm \frac{1}{3}\phi^\prime$, and
\begin{equation}\label{self}
\phi\left( y\right) =\left\{
\begin{array} {l l l}
\pm \frac{3}{4}\log\left| \frac{4}{3}y +c\right| +d & \mbox{for:}& y< 0 \\ 
\pm \frac{3}{4}\log\left| \frac{4}{3}y -c\right| +d & \mbox{for:}& y>0
\end{array}\right. ,
\end{equation}
where $d$ and $c$ are integration constants (they would correspond to 
the vacuum expectation values of moduli fields 
in an effective low energy description). Observe that with the
logarithm appearing in (\ref{self}) we are no longer dealing with
an exponential warp factor as (\ref{ansatz0}) would suggest. 
As a result of this we have to worry about possible singularities
in the solution under consideration. 
We shall come
back to this point in a moment.
Finally, by integrating the equations of motion around $y=0$ one
obtains the matching condition
\begin{equation}
T_0 = 4e^{\pm \frac{4}{3}d} .
\end{equation}
This means that the matching condition results in an adjustment of an
integration constant rather than a model parameter (like in the
previously discussed example). So, there seems to be no fine-tuning
involved even though we required $\bar{\Lambda} =0$. As long as one
can ensure that contributions to the vacuum energy on the brane couple
universely to the bulk scalar as given in (\ref{coupling}) it looks as
if one can adjust the vev of a modulus such that Poincare invariance
on the brane is not broken. 

In fact it seems that a miracle has
appeared: ``solution'' (\ref{self}) is apparently independent of the
brane tension $T_0$. So if one would add something to $T_0$
on the brane, the solution does not change. This would also solve
the problem of potential contributions to the brane tension
in perturbation theory, as they can be absorbed in $T_0$. Is 
this so-called self-tuning of the vacuum energy a solution to the
problem of the cosmological constant? Unfortunately not, since
there are some subtleties as we shall discuss now. 

We first notice that
the uniform coupling of the bulk scalar to
any contribution to the vacuum energy on the brane may be problematic
due to scaling anomalies in the theory living on the
brane \cite{Forste:2000ft}. 
Apart from that one would have to worry about the correct
strength of gravitational interactions.
In order to be in a agreement with four dimensional
gravity, the five dimensional gravitational wave equation should have
normalizable zero modes in the given background. In other words this
means that the effective four dimensional Planck mass should be
finite. 
For the model considered with a single brane at $y=0$
and $c < 0$ this implies
that $\int_{-\infty} ^{\infty} dy e^{2A\left( y\right)}  $
should be finite. However, plugging in the solution (\ref{self})
one finds that this is not satisfied. 
Following
\cite{Arkani-Hamed:2000eg,Kachru:2000hf} this could be solved 
by choosing $c>0$ and simultaneously 
cut off the $y$ integration at the
singularities at $\left|y\right| =\frac{3}{4} c$. This 
prescription then yields a
finite four dimensional Planck mass. 

With this choice of parameters, however, we are approaching 
disaster. Checking the
consistency condition (\ref{consistent}) one finds that it is not
satisfied anymore. The explanation for this is simple -- the equation
of motions are not satisfied at the singularities, and hence for
$c>0$  {\bf (\ref{self}) is  not a solution to the equations of motion}.  
It is the singularities that have created the miracle mentioned
above.

Of course, it has been often observed that singularities 
appear in an effective low energy prescription, and that at
those points new effects (such as massless particles) appear
as a result of an underlying theory to which the effective
description breaks down at this point. 
A celebrated example is  $N=2$
supersymmetric Yang-Mills theory where singularities in the moduli
space are due to monopoles or dyons becoming massless
at this point \cite{Seiberg:1994rs}. 
We might then hope that a similar mechanism (e.g. coming from
string theory) may
save the solution with $c>0$ and provide a solution to the problem of
the cosmological constant. It should be clear by now, that this
new physics at the singularity would be the solution of the
cosmological constant problem, if such a solution does exist at all.

In the following, we will investigate
such a mechanism and see whether it is connected to a potential 
fine-tuning of the parameters. Does the new physics 
at the singularity have to
know about the actual value of the tension of the brane
at $y=0$ or does it lead to a relaxation of the cosmological
constant independent of $T_0$?

To start this discussion, 
we first modify the theory in such a way that we obtain a
consistent solution in which the equations of motion are
satisfied everywhere. This can, for example, be done by adding two more
branes, situated at $\left| y\right|=\frac{3}{4} c$ to the setup. We then choose the
coupling of the bulk scalar to these branes as follows,
\begin{equation}
f_\pm \left(\phi\right)= T_{\pm}e^{b_\pm \phi} ,
\end{equation}
where the $\pm$ index refers to the brane at $y = \pm \frac{3}{4} c$. 
These two additional source terms in the action lead to two more
matching conditions whose solution is
\begin{equation}\label{finet}
b_+ =b_- = b = \mp\frac{4}{3}\,\,\,\,\,\, \mbox{and}\,\,\,\,\,\, T_+ =T_- =
-\frac{1}{2}T_0 . 
\end{equation}
It is obvious that here a third fine tuning (apart from $\Lambda_B=0$
and $b = \mp\frac{4}{3}$) has to be performed.
The amount of fine-tuning implied by these conditions is again
determined by the deviation of the vacuum energy on the brane from
the observed value. Hence, the situation has worsened
with respect to the question of fine-tuning. 
However, we have
learned that the consistency condition (\ref{consistent}) is a very
important tool to analyze the question of the cosmological constant.
A short calculation shows that (\ref{finet}) is essential for the
consistency condition (\ref{consistent}) to be satisfied.

\section{The moduli space of warped solutions}

So far, we focused on a very specific model and there remains the 
question whether this situation is generic. For the given set of
parameters we should then scan the available moduli space of
solutions parametrized by the values of the bulk cosmological
constant $\Lambda_B$, the brane tensions $T$ and the various
couplings like $b$ of the scalars to the brane. 
It is quite easy to see
that the above observation applies in general (for various explicit
examples see \cite{Forste:2000ft}). The reason is the fact that the 
amount of energy carried away from the brane by the bulk scalar needs
to be absorbed somewhere else. In principle, it could flow off to
infinity, but, as we have seen explicitely in the
last chapter, this cannot happen since in this case
we would not be able it to
localize gravity on the brane. In fact, it has been
shown in \cite{Csaki:2000wz} that localization of gravity is possible
only if there is either a fine-tuning between bulk and brane parameters 
as observed in the original Randall-Sundrum model or there are naked
singularities as in the models of
\cite{Arkani-Hamed:2000eg,Kachru:2000hf}. For the latter case 
the consistency condition (\ref{consistent}) requires the exactly 
fine tuned amount of energy from the singularity to match the
contribution from the branes.
We have seen this explicitly when studying a simple
way of ``resolving'' the singularities by adding new branes
with the appropriate tension. However, 
any other resolution of the singularities will lead to the same 
conclusions.  

So far we have concentrated on solutions that lead to flat
branes $\bar\Lambda=0$. A general discussion, however, should
also address the question whether the moduli space
of solutions also contains  ``nearby'' curved solutions that
are continuously connected to the flat solutions discussed so far. 
If they exist,
the solution of the cosmological constant problem would
need to supply arguments why the flat solutions are
favoured over these ``nearby'' curved solutions.

A first step in this direction would be to study
the response of the system once the
fine-tuning (which appears after the singularities are somehow
resolved) 
is relaxed. In various cases it has been shown 
that there exist also solutions with
$\bar{\Lambda} \not= 0$ \cite{Kachru:2000xs}.  
Moreover, for any fixed value of
$\bar{\Lambda}$  they fulfill 
the consistency condition (\ref{consistent})
for that value of $\bar{\Lambda}$
\cite{Forste:2000ft}. This means, that relaxing the fine-tuning 
to zero cosmological constant will lead to a consistent 
(curved) nearby solution
with a non-vanishing effective cosmological constant $\bar{\Lambda}$. 

It has been argued in the literature
that for the specific model which we discussed above
($b=\pm\frac{4}{3}$) there do not exist any nearby curved
solutions \cite{Arkani-Hamed:2000eg,Kachru:2000xs}. 
This would be a rather remarkable result 
as it would imply that in some way
the solution with $\bar\Lambda=0$ would be unique and
potentially stable. Observe that the option
of a smooth deformation of  $b$ away from $|b|=\frac{4}{3}$
is not possible since the $|b|\not=\frac{4}{3}$
solutions are not smoothly connected to the former
ones \cite{Kachru:2000hf}. The above argumention, however,
is only true under the asumption that the bulk cosmological
constant $\Lambda_B$ (or the bulk scalar potential) vanishes exactly.
The situation in which the scalar field received a
nontrivial bulk potential has been studied in \cite{Forste:2000ft}
with the result that depending on the value of the bulk potential
at zero the effective cosmological constant is constrained to a certain
non-zero value. 
Thus also the flat solution with $b=\pm\frac{4}{3}$ is
continuously connected to a nearby curved solution with 
$\bar\Lambda\not=0$ and nonvanishing bulk potential. In all the
known cases we thus see that the moduli space does not contain
isolated flat solutions. This is another way of stating the fact
that the problem of the cosmological constant has not been solved.

So far we have concentrated on the discussion of classical
solutions. As we have argued before there is a second 
aspect of the cosmological constant problem once we consider
quantum corrections as well. Generically we would asume that
radiative corrections would destroy any fine-tuning of the
classical theory if not forbidden by a symmetry. Supersymmetry
is an example, but since it is broken in nature at the TeV scale
it is not
sufficient to stabilize the vacuum energy to the degree
of say $10^{-3}$ eV. The special solution $|b|=\frac{4}{3}$
enjoys an additional symmetry, a variant of scale invariance.
This symmetry, however, has a quantum anomaly and therefore
cannot survive in the full quantum theory. A way out would be
to postulate a model with unbroken scale (or conformal)
invariance, i.e. a finite theory with vanishing 
$\beta$-function. But as in the case of  supersymmetry we know that this
symmetry cannot be valid far below the TeV region and thus 
cannot be relevant for the stability of the cosmological constant.

\section{Outlook}

From the above discussion it is clear that the brane world scenario 
gives a new view on the problem of the cosmological constant. 
However, the present discussion has not provided a
satisfactory solution, since in all the known cases 
a fine-tuning is needed to achieve agreement with
observations. In fact this fine-tuning is of the same order 
of magnitude as the one needed in ordinary four dimensional 
field theory. More work needs to be done to clarify the 
situation. One direction would be to analyze in detail the
possible implications of broken (bulk and brane) supersymmetry in the
general set-up. In the four-dimensional case we need
broken supersymmetry $M_{\rm SUSY}$ to be somewhere in the 
TeV region and also the value of the cosmological constant 
is essentially determined by $M_{\rm SUSY}$. In the
brane world scenario one could hope to separate these scales.
Some gymnastics in numerology would suggest 
$M_{\rm SUSY}^2/M_{\rm Planck}$ to be relevant for the 
(small but nonzero) size of the cosmological constant.
Unfortunately we have not yet found a satisfactory model
where such a relation is realized and the problem of the
size of the cosmological constant still has to wait for a
solution.

%\vfill\eject
\vskip 0.5cm
\noindent
{\large \bf Acknowledgements}

\smallskip
\noindent
It is a pleasure to thank Stefan F\"orste,
Zygmunt Lalak and St\'ephane Lavignac 
for collaboration on the subjects
presented in this talk. The help of Stefan F\"orste in the 
preparation of the present manuscript is highly appreciated.  
This work is supported by the European Commission RTN
programs HPRN-CT-2000-00131, 00148 and 00152.

%%%%%%%%%%%%%%%%%%%%%%
%%%%%%%%%%%%%%%%%%%%%%

\end{document}